\documentclass[aps,pre, preprintnumbers ,floats,twocolumn,superscriptaddress]{revtex4}
\usepackage{graphicx}
\usepackage{dcolumn}
\usepackage{bm}
\usepackage[english]{babel}
\usepackage[latin1]{inputenc}
\usepackage{graphicx}
\usepackage{epstopdf}
\usepackage{amsfonts}
\usepackage{color}
\usepackage{epsfig}
\usepackage{mathrsfs}
\usepackage{amsmath,amssymb}
\usepackage{subfigure}
\usepackage{multirow}
\usepackage{url}
\usepackage{dsfont}

\usepackage{framed}
\usepackage[svgnames]{xcolor}

\begin{document}

\selectlanguage{english}


\title{Evaluating the impact of interdisciplinary research: a multilayer network approach}

\author{Elisa Omodei, Manlio De Domenico and Alex Arenas\\Department of Mathematics and Computer Science, Rovira i Virgili University}


\begin{abstract}
Nowadays, scientific challenges usually require approaches that cross traditional boundaries between academic disciplines, driving many researchers towards interdisciplinarity. Despite its obvious importance, there is a lack of studies on how to quantify the influence of interdisciplinarity on the research impact, posing uncertainty in a proper evaluation for hiring and funding purposes.
Here we propose a method based on the analysis of bipartite interconnected multilayer networks of citations and disciplines, to assess scholars, institutions and countries interdisciplinary importance. 
Using data about physics publications and US patents, we show that our method allows to {\color{black} reward, using a quantitative approach, scholars and institutions that have carried out interdisciplinary work and have had an impact in different scientific areas.}
The proposed method could be used by funding agencies, universities and scientific policy decision makers for hiring and funding purposes, and to complement existing methods to rank universities and countries.
\end{abstract}

\maketitle

\flushbottom


\section{Introduction}

Interdisciplinary research has recently gained a central role in the advancement of science, leading to important achievements~\cite{nature_inter}. For instance, the 2014 Nobel Prize in Chemistry was awarded to two physicists and a physical chemist, for ``a physical technique, developed with help from chemistry, that helps illuminate problems in biology"~\footnote{An interdisciplinary celebration, \textit{Chemistry World} (2014) \protect\url{http://www.rsc.org/chemistryworld/2014/10/nobel-prize-editorial}}. 

Even though several definitions and metrics for interdisciplinarity have been proposed \cite{porter2007measuring,leydesdorff2007betweenness,wagner2011approaches,jensen2014many,sinatra2015century,lariviere2015long,UKreport}, citation impact metrics accounting for this aspect of scientific research have not been defined yet.

On the other hand, funding agencies have created specific calls for interdisciplinary projects, like the Interdisciplinary Programs funded by the National Science Foundation~\footnote{\protect\url{https://www.nsf.gov/od/iia/additional\_resources/interdisciplinary\_research/support.jsp}}.
The European Research Council explicitly encourages applications from scientists having published in multidisciplinary journals~\footnote{``Applicants should also be able to demonstrate a promising track-record of early achievements appropriate to their research field and career stage, including significant publications (as main author) in major international peer-reviewed multidisciplinary scientific journals, or in the leading international peer-reviewed journals of their respective field." \protect\url{http://erc.europa.eu/funding-and-grants/funding-schemes/starting-grants}}, and the evaluation criteria for the Marie Curie fellowships also include the interdisciplinary aspects of the research~\footnote{Annex 2 \protect\url{http://ec.europa.eu/research/participants/portal/doc/call/h2020/h2020-msca-if-2015/1645199-guide\_for\_applicants\_if\_2015\_en.pdf}}.
Consequently, there is significant need to evaluate projects and scholars by considering interdisciplinarity too.
The difficulties in evaluating interdisciplinary research constitute a pressing controversy that leads many young scholars to remain on more traditional tracks, because the risks associated to undertaking an interdisciplinary career path seem too high~\cite{rhoten2004risks}. 
This work addresses the issue of quantifying interdisciplinarity by proposing a method to rank scientific publications (such as papers and patents) and their producers (scholars, inventors, institutions, companies and countries) according to their scientific impact and its breadth over different scientific disciplines. The method is based on the detection of the most central elements of a complex bipartite interconnected multilayer network representing scientific producers and scientific citations within and across different fields. The citation network is composed of multiple layers, each representing a scientific discipline.  Accounting for this diversity -- instead of neglecting the information it provides by building an aggregated representation of the network -- allows to unveil the cross-disciplinary versatility of scientific publications and of their producers, and therefore to obtain a quantitative measure of their interdisciplinary scientific impact.

Since the seminal work of de Solla Price~\cite{de1965networks} and Garfield~\cite{garfield1979citation}, scientists have put a great effort into trying to understand the patterns of citation distributions~\cite{redner1998popular,king2004scientific,radicchi2008universality} and the non-trivial dynamics of scientific recognition~\cite{guimera2005team,newman2009first,eom2011characterizing,wang2013quantifying,penner2013predictability,zhang2013characterizing,uzzi2013atypical,deville2014career,ke2015defining}. This fundamental body of work has the ultimate goal of setting the basis for the definition of more accurate and fairer scientific impact metrics used for evaluation purposes.

In the last decades, several indices have been presented. They are based on the idea that we can quantify the impact of a scientific publication by counting the number of citations it has received over the years. 
A widely adopted indicator to evaluate scholars' scientific impact is the h-index~\cite{hirsch2005index} (a scholar has an index $h$ if $h$ of her/his publications have received at least $h$ citations each), and its numerous variants~\cite{egghe2006theory,bornmann2008there,kaur2013universality}.

More recently, a different approach has been proposed. Building networks that reconstruct the chains of scientific citations allows for a global understanding of the intrigued patterns of citations between publications -- or between producers. This representation allows to unveil the difference between, for instance, a publication that has received 10 citations coming from highly cited publications, and a publication that has received 10 citations too, but from low-cited papers. The two have the same number of citations but the former has clearly had a higher impact. To rank publications, journals or scholars according to their importance in the respective citation network, researchers have proposed diffusion algorithms that simulate the spreading of scientific credits on the network~\cite{walker2007ranking,bergstrom2007measuring,radicchi2009diffusion}. In practice, this is the same idea at the basis of the PageRank, i.e the algorithm that Google uses to rank the pages of the World Wide Web~\cite{brin1998anatomy}.

\section{Methodology}

\begin{figure}[!t]
\centering
\includegraphics[width=0.5\textwidth]{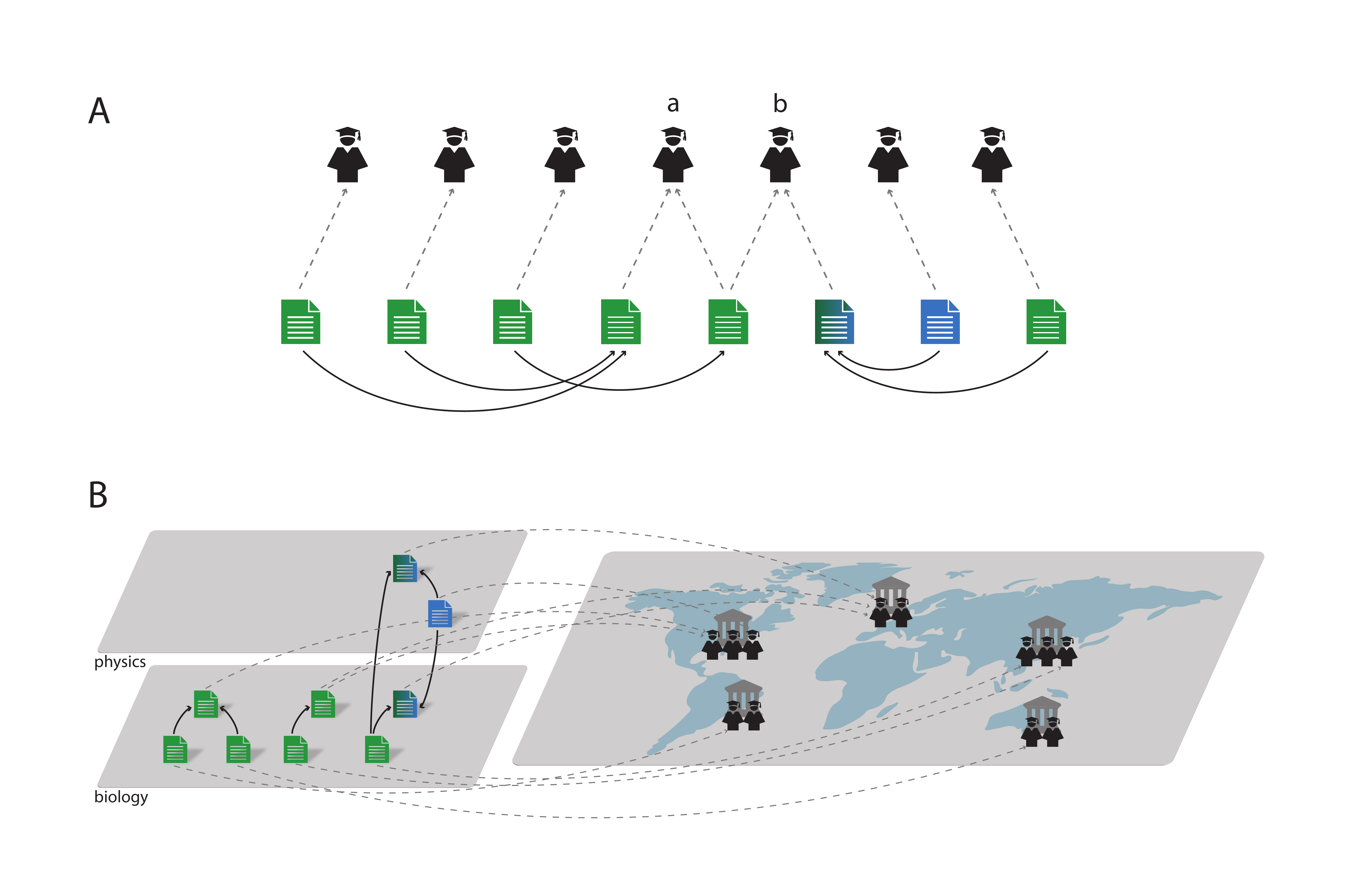}
\caption{\textbf{Bipartite interconnected multilayer network.} Panel A shows a simple example of bipartite citation network made of 8 papers and 7 scholars. The 8 papers belong to two disciplines -- biology and physics. Green icons represent biology papers, blue physics, and the bicolour icon represents a paper that belongs to both biology and physics. Continuous arrows represent citation edges, whereas dotted arrows connect papers to its authors. Panel B shows the multilayer representation of the network. Consider, for example, authors $a$ and $b$. If we discard the information about the scientific fields and consider the aggregated network shown in panel A, then the two authors' centrality would be the same, because they authored the same number of papers, having an identical structure of incoming citations. However, the multilayer framework takes into account that one of $b$'s papers pertains to both physics and biology, and, moreover, had an impact in both fields (one citation comes from a physics paper and the other from a biology one). Therefore $b$ has a higher versatility than $a$.}
\label{explicative-figure}
\end{figure}

In this work, we propose to rank scientific publications and their producers employing the PageRank defined on a bipartite interconnected multilayer structure that accounts for citations within and across different disciplines. This is equivalent to ranking nodes according to their \textit{versatility}~\cite{dedomenico2015} on an interconnected multilayer network~\cite{de2013mathematical,kivela2014multilayer}.

To account for interdisciplinarity, we define a bipartite interconnected multilayer network representing citations between publications (papers or patents) and relations between publications and their manufacturers (scholars, inventors, research institutions, companies or countries).

Given $N$ nodes and $L$ layers, the rank$-4$ multilayer adjacency tensor $A^{\alpha\tilde{\gamma}}_{\beta\tilde{\delta}}$ is defined in the following way. Let $C^{\alpha}_{\beta}(\tilde{h},\tilde{k})=\sum_{i,j=1}^{N}w_{i,j}(\tilde{h},\tilde{k})E^{\alpha}_{\beta}(ij)$ be the rank$-2$ adjacency tensor encoding information about the relationship between layer $\tilde{h}$ and $\tilde{k}$, where $w_{ij}(\tilde{h},\tilde{k})$ indicates the intensity of the relationship between node $n_i$ in layer $\tilde{h}$ and node $n_j$ in layer $\tilde{k}$, and $E^{\alpha}_{\beta}(ij)$ indicates the rank$-2$ tensor that represents the canonical basis in the space $\mathds{R}^{N\times N}$  (note that when $\tilde{h}=\tilde{k}$, $C^{\alpha}_{\beta}(\tilde{h},\tilde{h})$ represents the intra-layer adjacency tensor), then
\begin{equation}
A^{\alpha\tilde{\gamma}}_{\beta\tilde{\delta}}=\sum_{\tilde{h},\tilde{k}=1}^{L}C^{\alpha}_{\beta}(\tilde{h},\tilde{k})E^{\tilde{\gamma}}_{\tilde{\delta}}(\tilde{h},\tilde{k})
\end{equation}
where $E^{\tilde{\gamma}}_{\tilde{\delta}}(\tilde{h},\tilde{k})$ indicates the rank$-2$ tensor that represents the canonical basis in the space $\mathds{R}^{L\times L}$.
This is the general formulation of an adjacency tensor representing a multilayer network.

To build our network we consider $N=N_P+N_M$ nodes (where $N_P$ is the number of publications, and $N_M$ the number of manufacturers of the chosen type that produced the $N_P$ papers. Therefore, given the ordered set of nodes $\{n_1,...,n_N\}$, the first $N_P$ elements $\{ n_1,...,n_{N_P} \}$ represent publication, and the other $N_M$ elements $\{ n_{N_P+1},...,n_N \}$ represent manufacturers. Moreover, we consider $L=L'+1$ layers, where $L'$ is the number of scientific disciplines that the publications belong to. 
The 4 components of the rank$-2$ adjacency tensor $C^{\alpha}_{\beta}(\tilde{h},\tilde{k})$ are defined as follows.
$C^{\alpha}_{\beta}(\tilde{l_x},\tilde{l_x})$ and $C^{\alpha}_{\beta}(\tilde{l_x},\tilde{l_y})$, with $x,y \in [1,L']$, encode information about publication citations. Each layer represents a discipline or a subfield, therefore $w_{ij}(\tilde{l_x},\tilde{l_x}) = \frac{1}{N_L(i)N_L(j)}$ if both publications $i$ and $j$ belong to discipline $x$, and publication $i$ cites publication $j$. $N_L(i)$ ($N_L(j)$) is the number of disciplines that publication $i$ ($j$) belongs to. This normalisation is performed so that every citation carries one unit of value overall. Interdisciplinary citations are instead encoded by $C^{\alpha}_{\beta}(\tilde{l_x},\tilde{l_y})$; $w_{ij}(\tilde{l_x},\tilde{l_y}) = \frac{1}{N_L(i)N_L(j)}$ if publications $i$ belongs to discipline $x$ and $j$ to discipline $y$, and publication $i$ cites publication $j$. 
Let $\tilde{l_A}$ denote the remaining layer, then the tensors $C^{\alpha}_{\beta}(\tilde{l_x},\tilde{l_A})$, with $x \in [1,L']$, encode information about the relation between publications and their manufacturers, \textit{i.e.} if the chosen type of manufacturer is scholars, then $w_{ij}(\tilde{l_x},\tilde{l_A}) = \frac{1}{N_L(i)}$ if author $j$ is one of the authors of publication $i$. If we consider research institutions, we connect each publication to the institutions to which its authors are affiliated; if we consider countries, the connections are to the countries in which these institutions are based. Finally, we define $C^{\alpha}_{\beta}(\tilde{l_A},\tilde{l_x})$ and $C^{\alpha}_{\beta}(\tilde{l_A},\tilde{l_A})$ to be zero tensors. $C^{\alpha}_{\beta}(\tilde{l_A},\tilde{l_x})$ tensors are null because we do not want the relations between publications and manufacturers to be symmetric, to avoid unrealistic paths to take place when computing the nodes centrality. $C^{\alpha}_{\beta}(\tilde{l_A},\tilde{l_A})$ is null because all the information is already encoded in the other tensors: we do not need to explicitly add citation edges between authors.

In this framework, for the citation layers, each node is active on a given layer if and only if the publication it represents belongs to the corresponding field. For example, a monodisciplinary publication is active only on one layer, whereas an interdisciplinary publication, pertaining to both physics and biology, is active on two layers. As a consequence, a publication whose impact is restricted to only one discipline has intra-layer incoming edges only, whereas a publication that has influenced the work of researchers in more than one field has inter-layer incoming edges too, which represent the bridges between the different fields involved. Therefore this framework allows for a natural representation of the interdisciplinarity degree of a publication. Being our goal to rank publication producers too, we introduce in the network a second type of nodes, which, according to the specific need, represent scholars, inventors, research institutions, or countries. These nodes are active on a dedicated layer, and each publication has directed outgoing inter-layer edges pointing to each of its producers. 
Previous works on ranking producers are based on one-mode projections of the bipartite network of publications and producers, whereas in this work we prefer to take advantage of the complete bipartite structure in order to avoid any information loss, as further detailed in the Supplementary Material. 

On the proposed network, the ranking is obtained through a process of diffusion of scientific credits from paper to paper through citation edges within and across disciplines.
Producers are the sinks of this diffusion process, being represented by nodes with no outgoing edges, and incoming edges originated from the papers they have produced.
A schematic representation of the proposed network is shown in Figure~\ref{explicative-figure}.

Having defined the multilayer citation network, we propose to rank its nodes according to their PageRank versatility, which is given by the steady-state solution of the equation
\begin{equation}
p_{\beta\tilde{\delta}}(t+1) = R^{\alpha\tilde{\gamma}}_{\beta\tilde{\delta}} p_{\alpha\tilde{\gamma}}(t)
\end{equation}
where $p_{\alpha\tilde{\gamma}}(t)$ is the time-dependent tensor that gives the probability to find a random walker at a particular node $\alpha$ in a particular layer $\tilde{\gamma}$, and
\begin{equation}
R^{\alpha\tilde{\gamma}}_{\beta\tilde{\delta}} = \Big[ r T^{\alpha\tilde{\gamma}}_{\beta\tilde{\delta}} + \frac{1-r}{NL} u^{\alpha\tilde{\gamma}}_{\beta\tilde{\delta}} \Big] \mbox{,}
\end{equation}
$N$ being the number of nodes in the network, $L$ the number of layers, and $r$ the teleportation rate. $T^{\alpha\tilde{\gamma}}_{\beta\tilde{\delta}}$ denotes the rank$-4$ tensor of transition probabilities for jumping between pairs of nodes and switching between pairs of layers, and $u^{\alpha\tilde{\gamma}}_{\beta\tilde{\delta}}$ is the rank$-4$ tensor with all components equal to 1. 
Let $\Omega_{\alpha\tilde{\gamma}}$ be the eigentensor of the transition tensor $R^{\alpha\tilde{\gamma}}_{\beta\tilde{\delta}}$, denoting the steady-state probability to find a random walker in node $\alpha$ and layer $\tilde{\gamma}$. The tensor $\Omega_{\alpha\tilde{\gamma}}$ provides the PageRank of each node ($\alpha$) in each layer ($\tilde{\gamma}$): it is crucial to remark here that this is not equivalent to calculate the PageRank in each layer separately, because our formulation accounts for the whole interconnected structure to solve the eigenvalue problem. To obtain the multilayer PageRank of each node, regardless of the layer, we project the values obtained from its replicas in different layers, obtaining the multilayer PageRank vector
\begin{equation}
\omega_{\alpha} = \Omega_{\alpha\tilde{\gamma}} u^{\tilde{\gamma}}
\end{equation}
where $u^{\tilde{\gamma}}$ is the vector with all components equal to 1. It has been shown~\cite{dedomenico2015} that this operation provides the same results that would be obtained by calculating PageRank by means of simulated random walkers that explore the multilayer structure according to transition rules encoded in $R^{\alpha\tilde{\gamma}}_{\beta\tilde{\delta}}$.

\section{Data}

To illustrate the proposed ranking method, we test it on two case studies: the American Physical Society (APS) and the US patents datasets.

The first is a collection of papers published in the journals of the American Physical Society (Physical Review Letters, Physical Review and Review of Modern Physics) between 1985 and 2009~\footnote{Data provided by APS upon request, \protect\url{https://publish.aps.org/datasets}}. 
We restricted the analysis only to papers with at most ten authors, to avoid biases due to the papers of experimental high-energy physics in which all the project collaborators are listed as co-authors. To disambiguate author's name, we used a simple technique introduced in previous studies~\cite{radicchi2009diffusion}.
Meta-data in the dataset provide information about the topic of the papers through the specification of the assigned ``Physics and Astronomy Classification Scheme'' (PACS) code, developed by the American Institute of Physics (AIP) and used in Physical Review since 1975 to identify fields and sub-fields of physics~\footnote{\protect\url{http://journals.aps.org/PACS}}.  
We exploited this information to build a heterogeneous interconnected 10-layer network in which each layer represents a sub-field of physics, as defined by the PACS systems: \textit{General}; \textit{The Physics of Elementary Particles and Fields}; \textit{Nuclear Physics}, \textit{Atomic and Molecular Physics}; \textit{Electromagnetism, Optics, Acoustics, Heat Transfer, Classical Mechanics, and Fluid Dynamics}; \textit{Physics of Gases, Plasmas, and Electric Discharges}; \textit{Condensed Matter: Structural, Mechanical and Thermal Properties}; \textit{Condensed Matter: Electronic Structure, Electrical, Magnetic, and Optical Properties}; \textit{Interdisciplinary Physics and Related Areas of Science and Technology}; \textit{Geophysics, Astronomy, and Astrophysics}.
From the paper meta-data we also extracted the authors affiliation information, which allowed us to associate to each paper a list of (one or more) institutions and countries. The final dataset consists of 319816 papers, 204809 authors, 626 institutions and 54 countries.
Arguably, the APS dataset covers only Physics, but note that physics is a vast field that spans from biological physics to astrophysics and although it may fall short of a full interdisciplinary analysis it is clear that this is a powerful indicator of multi-topic analysis that serves to proof the usefulness of the method.

The second dataset contains the U.S. patents granted between January 1963 and December 1999, and all citations made to these patents between 1975 and 1999~\footnote{\protect\url{http://www.nber.org/patents/}}. 
To define the layers, we used the 6 categories proposed in previous studies~\cite{hall2001nber}: \textit{Chemical (excluding Drugs)}; \textit{Computers and Communications}; \textit{Drugs and Medical}; \textit{Electrical and Electronics}; \textit{Mechanical}; \textit{Others}.
Each patent is assigned to one main class defined by the United States Patent and Trademark Office (USPTO), and to any number of subsidiary classes. Each class belongs to one of the listed categories, therefore each patent is associated with one or more layer according to its classes. 
However, the dataset only contains the information about the main class, therefore we complemented it by extracting the information about the other classes from the USPTO Patent Grant Full Text~\footnote{\protect\url{http://www.google.com/googlebooks/uspto-patents-grants-text.html}}.
The final dataset contains 1574882 patents, 1142499 inventors, 138833 assignees (i.e. corporations for the most part), and 127 countries.

\section{Results}

Figure~2 shows the evolution of the interdisciplinary ranking of the world top physics departments, and of the world top companies, over time. This visualization allows to observe, for instance, the raise of the University of Texas at Austin during the 1990s, after the establishment, in 1985, of the Center for Nonlinear Dynamics, funded and directed by the Boltzmann Medal laureate Harry Swinney~\footnote{\protect\url{https://web2.ph.utexas.edu/utphysicshistory/UTexas\_Physics\_History/Center\_for\_Nonlinear\_Dynamics.html}}.

\begin{figure}[!t]
\centering
\includegraphics[width=0.5\textwidth]{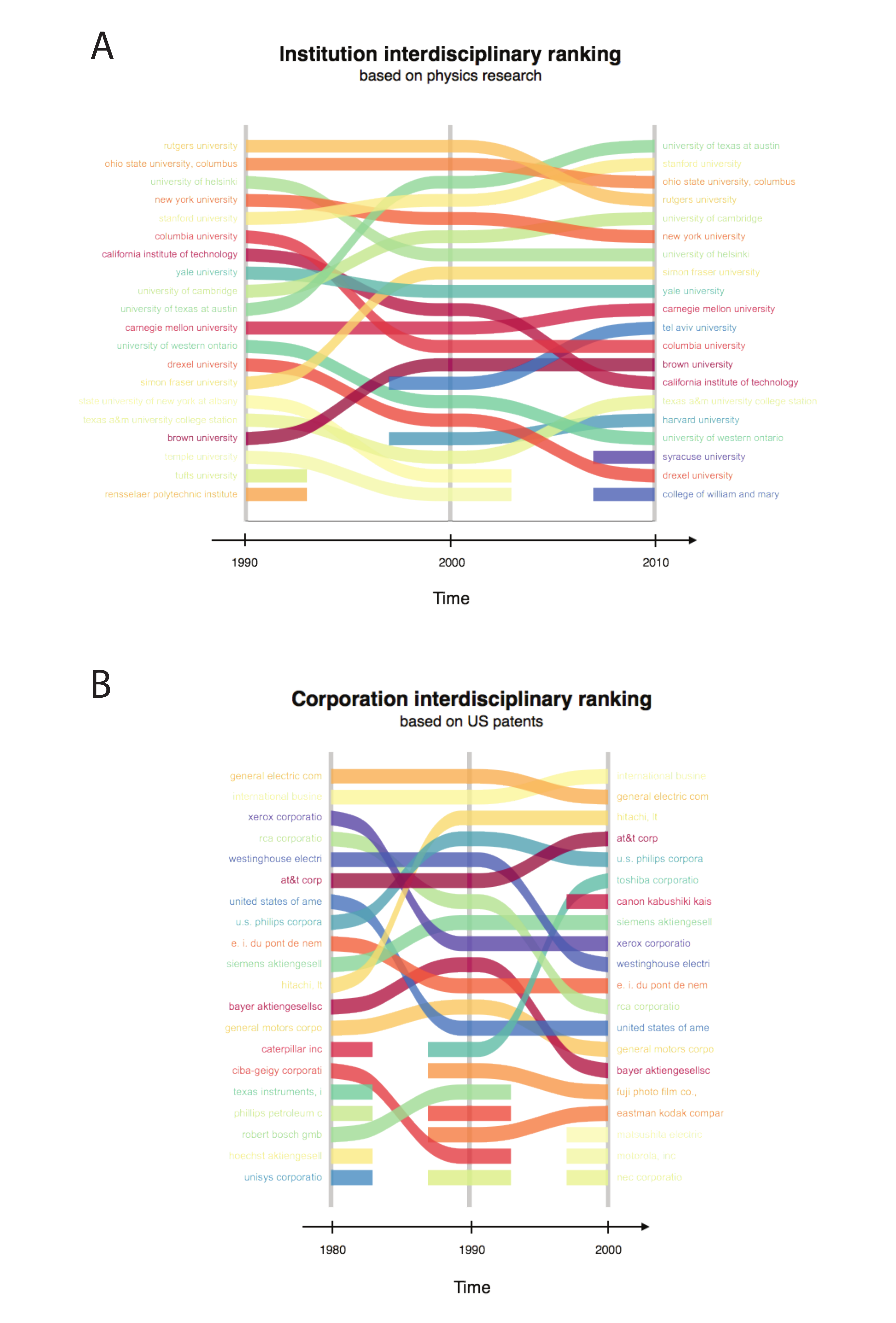}
\caption{\textbf{Interdisciplinary ranking evolution.} Panel A: Visualization of the time evolution of the interdisciplinary impact ranking of the top 20 physics departments, computed using the APS dataset. {\color{black}The rank is visualized top-down, i.e. the top institution is the first ranked.} Panel B: Time evolution of the interdisciplinary impact ranking of the top 20 world companies, computed using the US patent data. Broken lines represent institutions or companies that do not belong to the top 20 in the previous or the following time stamp.}
\label{rankings}
\end{figure}

{\color{black} Compared to previously proposed algorithms of diffusion of scientific credit, the proposed method rewards researchers that have carried out interdisciplinary works or have had an impact in different scientific areas. To show this, we first compare it with the science author rank algorithm (SARA)~\cite{radicchi2009diffusion}. We find that the rankings of APS authors obtained using SARA and using the proposed method have a Spearman's rank correlation of 0.77 ($99\%$ confidence level). The high value of correlation is to be expected since both methods rank researchers simulating a diffusion process on a citation network. However, the proposed method gives higher ranking to versatile researchers such as the self-organised criticality pioneer Per Bak (+21 positions gained), econophysics co-founder Eugene Stanley (+56), complex networks pioneer Shlomo Havlin (+104), and complex systems professor Leonard M. Sander (+93).}

We show that the proposed method is in fact able to capture two fundamental aspects of interdisciplinary research: intrinsic multidisciplinarity (i.e. publishing papers or patents pertaining to different areas) on the one hand, and effective interdisciplinarity, i.e. being credited by different scientific areas, on the other.

We define the \textit{topical interdisciplinarity} $TI(a)$ of an author $a$ (who could be a scholar or an inventor) as the average number of different scientific areas her/his publications pertain to, i.e.:
\begin{equation}
TI(a) = \frac{1}{n(a)} \sum_{i=1}^{n(a)} d(p_i)
\end{equation}
where $n(a)$ is the number of publications authored by $a$, and $d(p_i)$ is the number of fields that publication $p_i$ belongs to.

Moreover, for each publication $p$ we define an entropy metrics based on the distribution of its incoming citations across the different fields represented by the different layers: 
\begin{equation}
H(p) = \sum f_i \log{\frac{1}{f_i}}
\end{equation}
where the sum is over the different fields (layers) and $f_i$ is the proportion of edges incident in $p$ that come from layer $i$. Therefore if a publication is only cited by other publications belonging to its own field $H(p)=0$, whereas a publication that has received citations from different fields has $H(p)>0$, and the higher the number of fields it has had impact on, the higher its entropy.
For each author $a$ we then compute her/his \textit{citation interdisciplinarity} $CI(a)$ as the average entropy of her/his publications:
\begin{equation}
CI(a) = \frac{1}{n(a)} \sum_{i=1}^{n(a)} H(p_i) \mbox{.}
\end{equation}

We find a strong positive correlation between the gain in rank that scholars and inventors obtain when evaluated using the proposed method -- instead of a method based on a flat representation of the citation network --, and their topical interdisciplinarity (Figure~3, panels (a) and (b)). Moreover, we find that the rank gain is also positively correlated with the disciplinary diversity of scholars' and inventors' incoming citations (Figure~3, panels (c) and (d)). {\color{black} To control for the effects of productivity, i.e. the fact that a researcher that produces more papers has more chances to publish in more areas or to be cited by papers in many different areas, we perform the same analysis on two subsets of the data by considering in each case only authors with a fixed number of publications ($20\pm2$ and $50\pm2$). The correlation coefficients found in these subsets are consistent with those found using the whole dataset, demonstrating that the proposed method is not biased by productivity. The results are reported in the Supplementary Material.}

\begin{figure}[!t]
\centering
\includegraphics[width=0.5\textwidth]{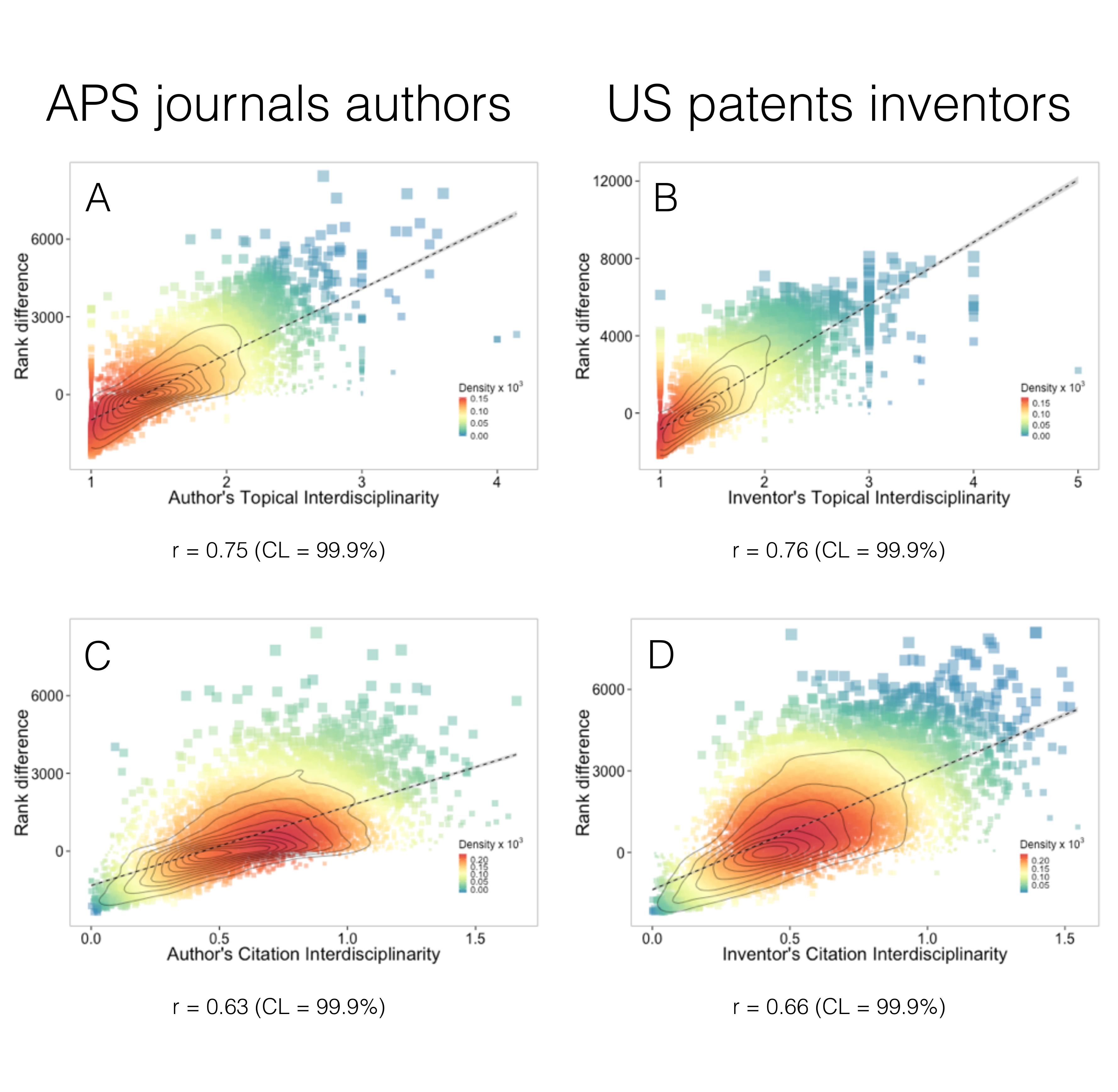}
\caption{\textbf{Correlations.} Heat-maps representing the correlation between the gain in rank that scholars and inventors obtain when evaluated using the proposed method -- instead of a method based on a flat representation of the citation network --, and two measures of their interdisciplinarity level. The x-axis represents, in panel (a) and (b), scholars' and inventors' topical interdisciplinarity, defined as the average number of different scientific areas their publications pertain to, and, in panel (c) and (d), their diversity in terms of disciplines of the scholars' and inventors' incoming citations (citation interdisciplinarity). Correlations are calculated using Pearson's $r$ coefficient, and setting the statistical significance at $0.1\%$. Solid lines represent density gradient contours, and dashed lines represent linear regression models estimated via maximum-likelihood.}
\label{correlations}
\end{figure}

\section{Discussion}
In this paper, we propose a methodology to assess the citation impact of scientific publications and their producers that intrinsically accounts for their interdisciplinarity. This aspect was not included in previous citation impact indicators.

Even though numerous metrics have been proposed to assess citation impact, several issues have been raised. These include the accounting of self-citations~\cite{glanzel2006concise}, the choice of the appropriate citation time window~\cite{wang2013citation}, field normalisation~\cite{li2013quantitative}, and author credit allocation~\cite{gauffriau2008comparisons}. Despite the vast literature on the subject, consensus is still lacking on how to solve these issues. Here, we propose a method whose objective is to account for interdisciplinarity, and we do not enter those debates. However, the bipartite interconnected multilayer networks of citations and disciplines that we introduce can be adapted to take into account specific needs. For example, edges connecting papers to their authors could be weighted differently to take into account non-homogeneous allocation of credit, or a specific time window could be chosen a priori to select the papers constituting the network.

Going beyond the presented assessment of the benefits produced by interdisciplinarity, the method proposed in this work could constitute a tool for funding agencies and academic hiring decision makers to quantify the impact of interdisciplinary research and its producers, for a faster advancement of excellent science. 

\section*{Acknowledgements}
A.A., E.O. and M.D.D. acknowledge financial support by the Spanish government through grant FIS2015-38266. M.D.D. also acknowledges financial support from the Spanish program Juan de la Cierva (IJCI-2014-20225). A.A. also acknowledges partial financial support from the European Commission FET-Proactive project MULTIPLEX (Grant No. 317532), ICREA Academia and James S.\ McDonnell Foundation.
\small

\appendix

\section{Comparison with other approaches}

The idea of ranking scholars simulating a diffusion process was already introduced in \cite{radicchi2009diffusion} -- and previously in \cite{walker2007ranking} to rank scientific publications -- but the approach proposed in this work considers a different kind of network -- a bipartite interconnected multilayer network. In this section we motivate the choice of taking into account the complete bipartite structure instead of its one-mode projection.

For the sake of simplicity, we will consider a bilayer version of the network. Our focus here is not in fact the multilayer aspect of the network capturing interdisciplinarity, but rather its bipartition. Therefore, let us consider $N=N_P+N_A$ nodes and 2 layers $\{l_1,l_2\}$. 
The 4 components of the rank$-2$ adjacency tensor $C^{\alpha}_{\beta}(\tilde{h},\tilde{k})$ are now defined as follows. $C^{\alpha}_{\beta}(\tilde{l_1},\tilde{l_1})$ encodes information about citing relations between papers, \textit{i.e.} $w_{ij}(\tilde{l_1},\tilde{l_1}) = 1$ if paper $i$ cites paper $j$. $C^{\alpha}_{\beta}(\tilde{l_1},\tilde{l_2})$ encodes information about paper authorship, \textit{i.e.} $w_{ij}(\tilde{l_1},\tilde{l_2}) = 1$ if author $j$ is one of the authors of paper $i$. Finally, we define $C^{\alpha}_{\beta}(\tilde{l_2},\tilde{l_1})$ and $C^{\alpha}_{\beta}(\tilde{l_2},\tilde{l_2})$ to be zero tensors, consistently with the representation introduced in S.1. For the sake of simplicity, since in the rest of the section we will be dealing only with rank-2 tensors, we will make use of the simpler classical matrix notation instead of the tensorial one. Therefore we will denote $C^{\alpha}_{\beta}(\tilde{l_1},\tilde{l_1})$ as $C$ and $C^{\alpha}_{\beta}(\tilde{l_1},\tilde{l_2})$ as $A$.

In the author citation network proposed in \cite{radicchi2009diffusion}, each node represents an author, and $w_{ij} \neq 0$ if there exist at least one publication $\alpha$, of which $i$ is an author, that cites a publication $\beta$ of which $j$ is an author. Each such publication gives a contribution $\frac{1}{nm}$ to $w_{ij}$ (where $n$ is the number of authors of publication $\alpha$, and $m$ is the number of authors of $\beta$) so that the total contribution of each citation is equal to 1.

Let us consider an adjacency matrix $\mathcal{C}$ of size $N_P \times N_P$, encoding the citation links between papers. $C$ can be built from $\mathcal{C}$ by means of multiplication with a rectangular matrix $\mathcal{I}$ of size $(N_P+N_A) \times N_P$ such that $(\mathcal{I})_{ii} = 1$ for $i=1,...,N_P$, and all the other elements are equal to 0. Then
\begin{equation}
C = \mathcal{I} \mathcal{C} \mathcal{I}^T \mbox{.}
\end{equation}
Using $\mathcal{I}$ we can also build $A$ from the projection matrix $\mathcal{P}$, of size $N_P \times N_A$, where $w_{ij} = 1$ if $j$ is one of the authors of paper $i$:
\begin{equation}
A = \mathcal{I} \mathcal{P} \mathcal{I}^T \mbox{.}
\end{equation}
Let us define $\tilde{\mathcal{P}}$ as the normalised version of $\mathcal{P}$, \textit{i.e.} $(\tilde{\mathcal{P}})_{ij} = \frac{(\mathcal{P})_{ij}}{\sum^{N_A}_{k=1}(\mathcal{P})_{ik}} = \frac{1}{m_i}$ (where $m_i$ is the number of authors of paper $i$), then the $N_A \times N_A$ adjacency matrix representing the network of citations between authors can be obtained performing two successive matrix multiplications:
\begin{equation}
\mathcal{A} = \tilde{\mathcal{P}}^T \mathcal{C} \tilde{\mathcal{P}} \mbox{.}
\end{equation}
\textit{Proof:}
\begin{equation}
(\tilde{\mathcal{P}}^T \mathcal{C})_{ik} = \sum^{N_P}_{h=1} (\mathcal{P})_{hi} (\mathcal{C})_{hk} = \sum^{N_P}_{\substack{h=1 \\(\mathcal{P})_{hi} \neq 0, (\mathcal{C})_{hk} = 1}} \frac{1}{m_h}
\end{equation}
This means that $(\tilde{\mathcal{P}}^T \mathcal{C})_{ik}$ is a sum over the papers authored by $i$ that cite paper $k$, where each paper $h$ gives a contribution of 1 over the number of authors. Then:
\begin{equation}
(\mathcal{A})_{ij} = \sum^{N_P}_{k=1} (\tilde{\mathcal{P}}^T \mathcal{C})_{ik} (\mathcal{P})_{kj} = \sum^{N_P}_{\substack{k=1 \\(\mathcal{P})_{kj} \neq 0}} \Big( \sum^{N_P}_{\substack{h=1 \\(\mathcal{P})_{hi} \neq 0, (\mathcal{C})_{hk} = 1}} \frac{1}{m_h} \Big) \frac{1}{m_k}
\end{equation}
Each element $(\mathcal{A})_{ij}$ is therefore a sum over all the pairs of papers $(h,k)$ such that $i$ is an author of $h$ and $j$ of $k$, and each element of the sum gives a contribution equal to $\frac{1}{m_hm_k}$, as indeed defined in \cite{radicchi2009diffusion}. $\square$

\begin{figure}
\centering
\includegraphics[width=0.5\textwidth]{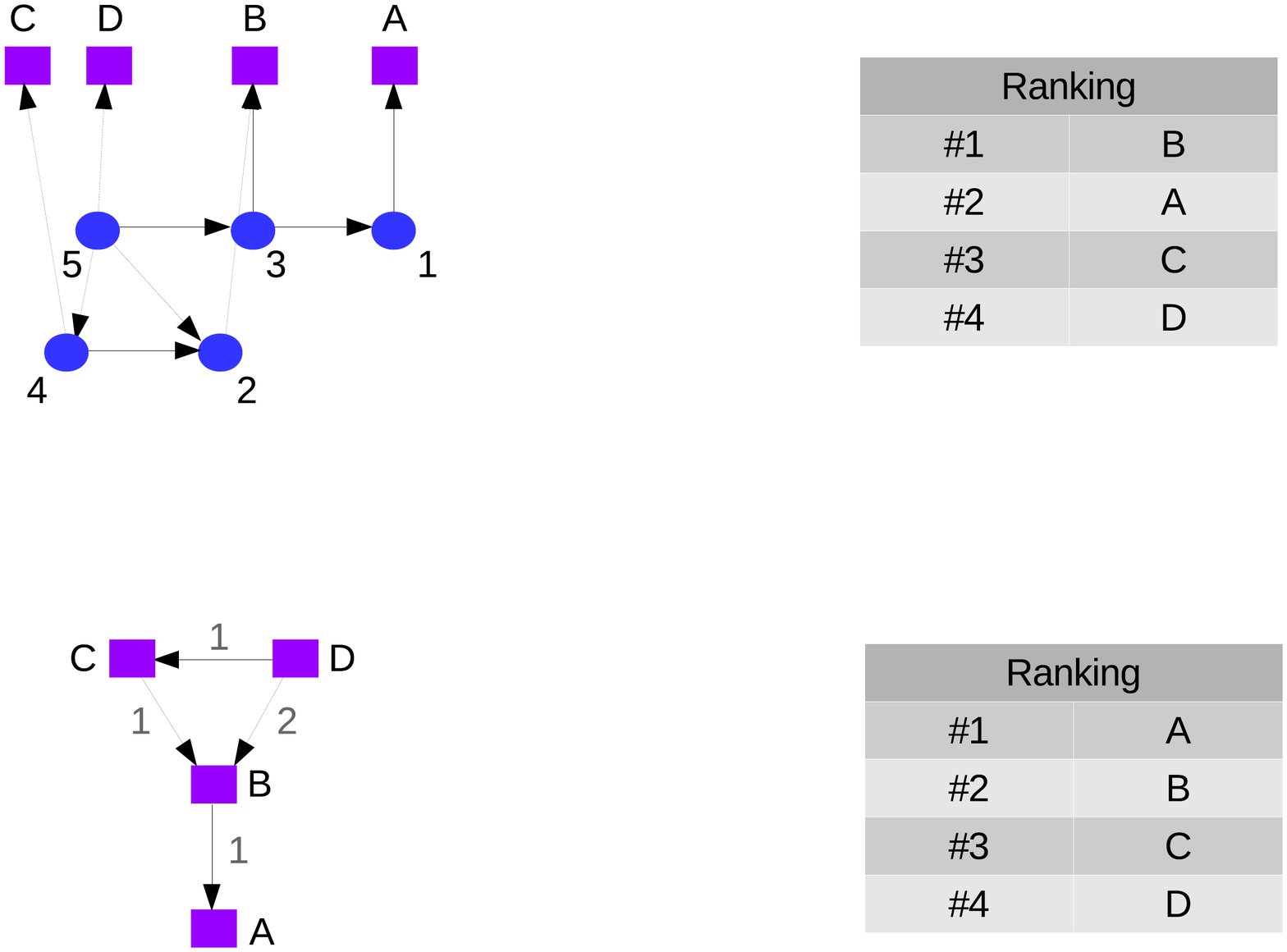}
\caption{An example in which the PageRank centralities computed on the author citation network and on the bipartite network lead to different rankings of the authors.}
\label{toy-model-1}
\end{figure}

We have demonstrated that the adjacency matrix of the author citation network is obtained performing two operations of matrix multiplication involving $\mathcal{C}$ and $\mathcal{P}$. Matrix multiplications consists in multiplications and summations of the matrix elements, which inevitably lead to information loss. The supra-adjacency matrix of the network proposed in this paper is instead the sum of the two expansions of $\mathcal{C}$ and $\mathcal{P}$, \textit{i.e.} $C$ and $A$, respectively. This guarantees that no information is loss, and this why we chose to consider the whole bipartite structure. Figure \ref{toy-model-1} shows an example in which the information loss characteristic of the author citation network leads to a less fair ranking of authors compared to the ranking based on the network introduced in this paper. Using our approach (top figure), the most central author is $B$, who is the author of both the most central paper and of one of the second most central papers. However, in the author citation network framework, the most central author is $A$ (to understand why, we recall that the PageRank centrality is based not only on the number of incoming edges, but on the importance of the nodes from which these edges originate). This is due to the fact that in the author citation network all the information about an author's incoming citations from different papers is aggregated, and therefore $A$ benefits from the importance of $B$ without any distinction between the importance coming from papers that actually cite $A$, and that coming from $B$'s other papers. In this case $B$'s most central (and cited) paper is not the one citing $A$'s paper, but this information is lost in the author citation network. As a consequence, the resulting ranking does not always reflect the real importance of the different authors.

\begin{figure}
\centering
\includegraphics[width=0.5\textwidth]{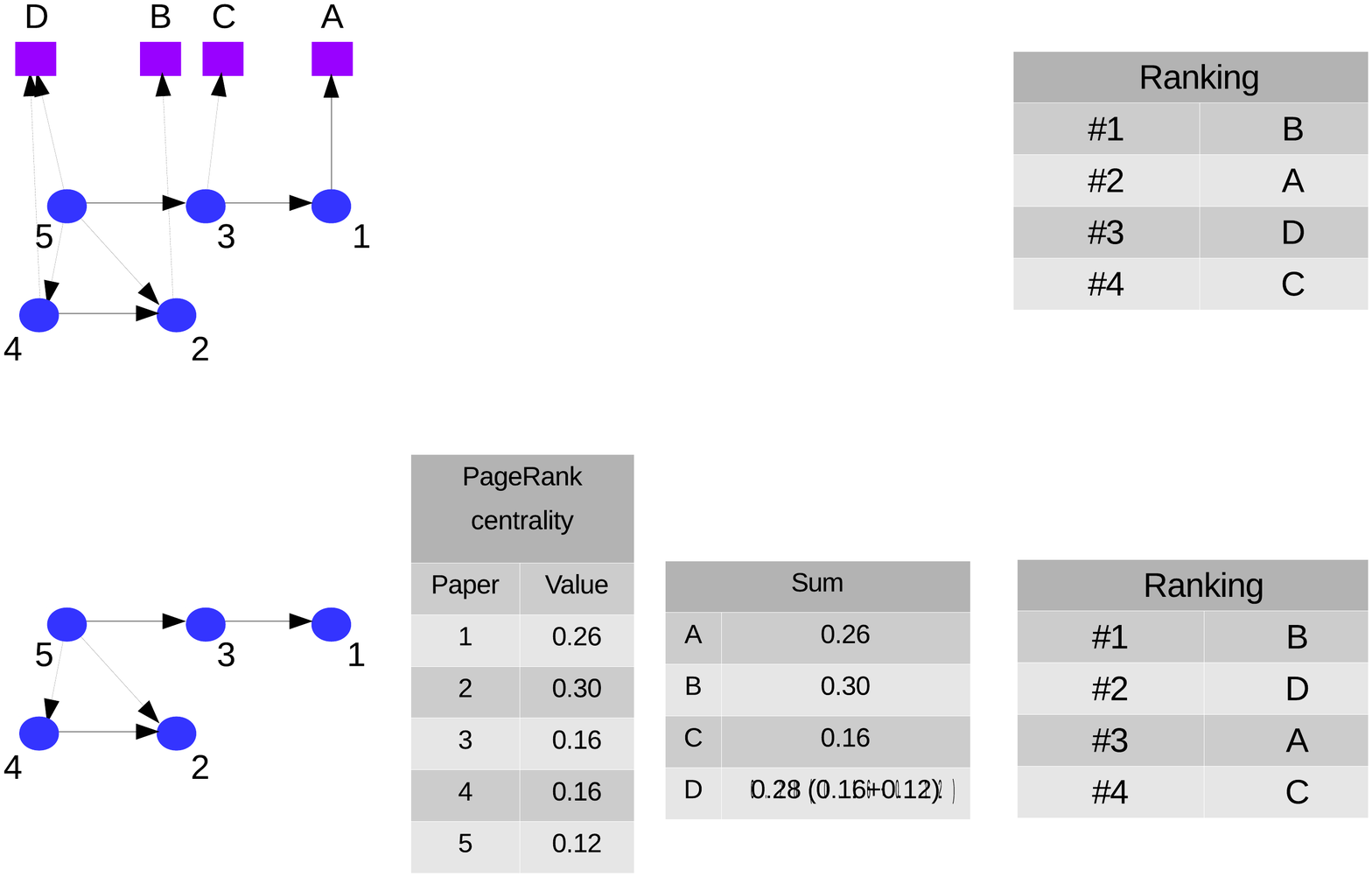}
\caption{An example in which the PageRank centralities computed as the sum over the papers and on the bipartite network lead to different rankings of the authors.}
\label{toy-model-2}
\end{figure}

An alternative approach is to use the PageRank method to get the centrality of papers in $\mathcal{C}$, and then compute the author centrality as a properly normalised sum of the centralities of the papers she/he has authored. In matrix terms, the author PageRank centrality vector $\mathbf{\omega}_A$ can be obtained by simply applying a linear transformation to the paper PageRank centrality vector $\mathbf{\omega}_P$:
\begin{equation}
\omega_A = \tilde{\mathcal{P}}^T \omega_P \textit{.}
\end{equation}
However, this solution involves another kind of aggregation which can lead to misleading results too. An example is shown in Figure \ref{toy-model-2}. Using the sum approach, author $D$ becomes more central than author $A$, because she/he authored two papers, even though they are the two most marginal papers in the network (note that the only citation to paper 4 is a self-citation). On the contrary, $A$ is the author of a very central paper, and in fact our approach correctly classifies her/him as more central than $D$. The issue with this alternative approach is that the PageRank is a diffusion process, which is not a linear dynamics. Therefore summing over the centralities of different nodes is also an aggregation process through which some information on the system is lost.

\section{Productivity control}

\begin{figure}
\centering
\includegraphics[width=0.5\textwidth]{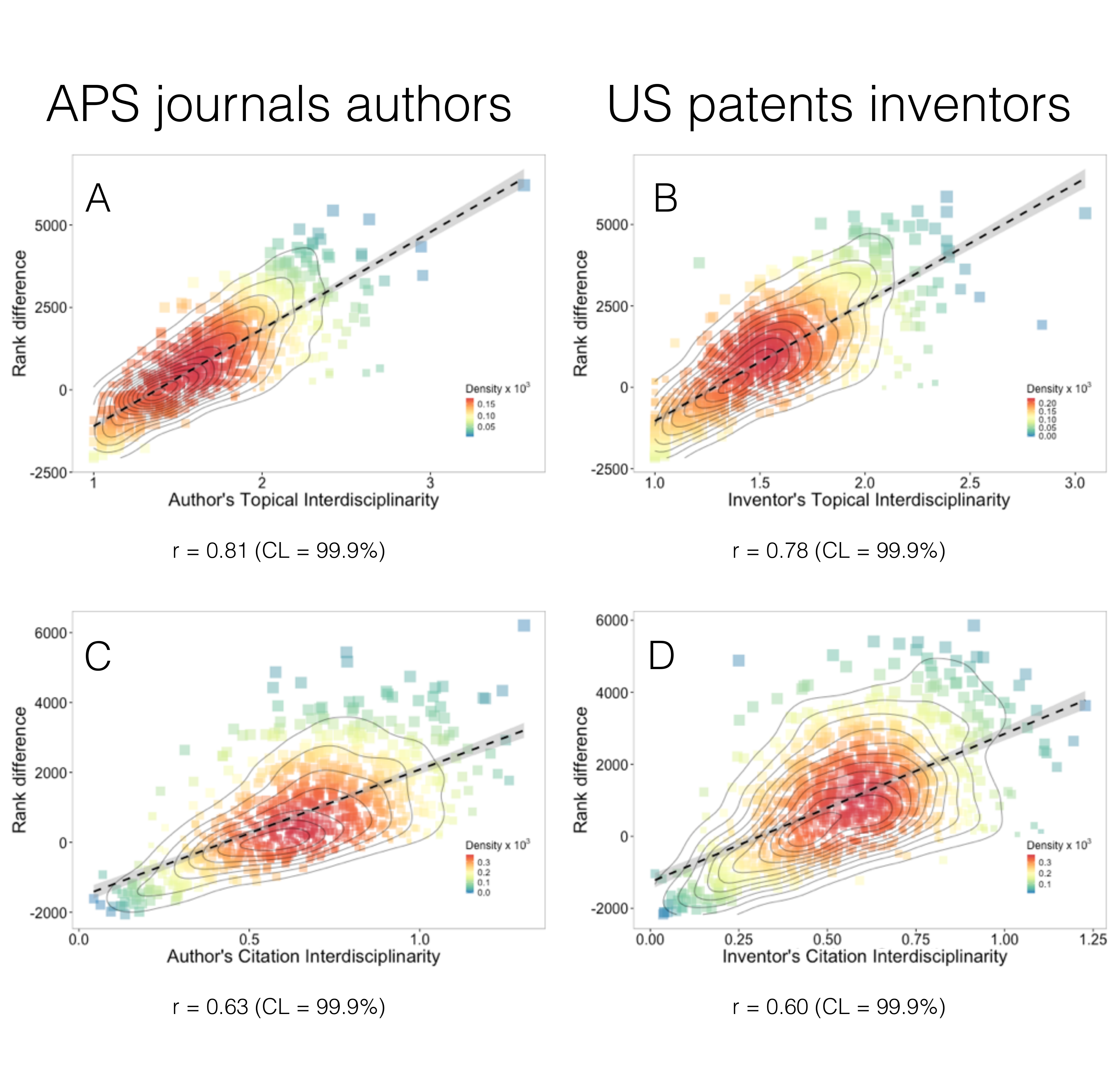}
\caption{\textbf{Correlations.} Heat-maps representing the correlation between the gain in rank that scholars and inventors with $20\pm2$ publications obtain when evaluated using the proposed method -- instead of a method based on a flat representation of the citation network --, and two measures of their interdisciplinarity level. The x-axis represents, in panel (a) and (b), scholars' and inventors' topical interdisciplinarity, defined as the average number of different scientific areas their publications pertain to, and, in panel (c) and (d), their diversity in terms of disciplines of the scholars' and inventors' incoming citations (citation interdisciplinarity). Correlations are calculated using Pearson's $r$ coefficient, and setting the statistical significance at $0.1\%$. Solid lines represent density gradient contours, and dashed lines represent linear regression models estimated via maximum-likelihood.}
\label{control_20papers}
\end{figure}

\begin{figure}
\centering
\includegraphics[width=0.5\textwidth]{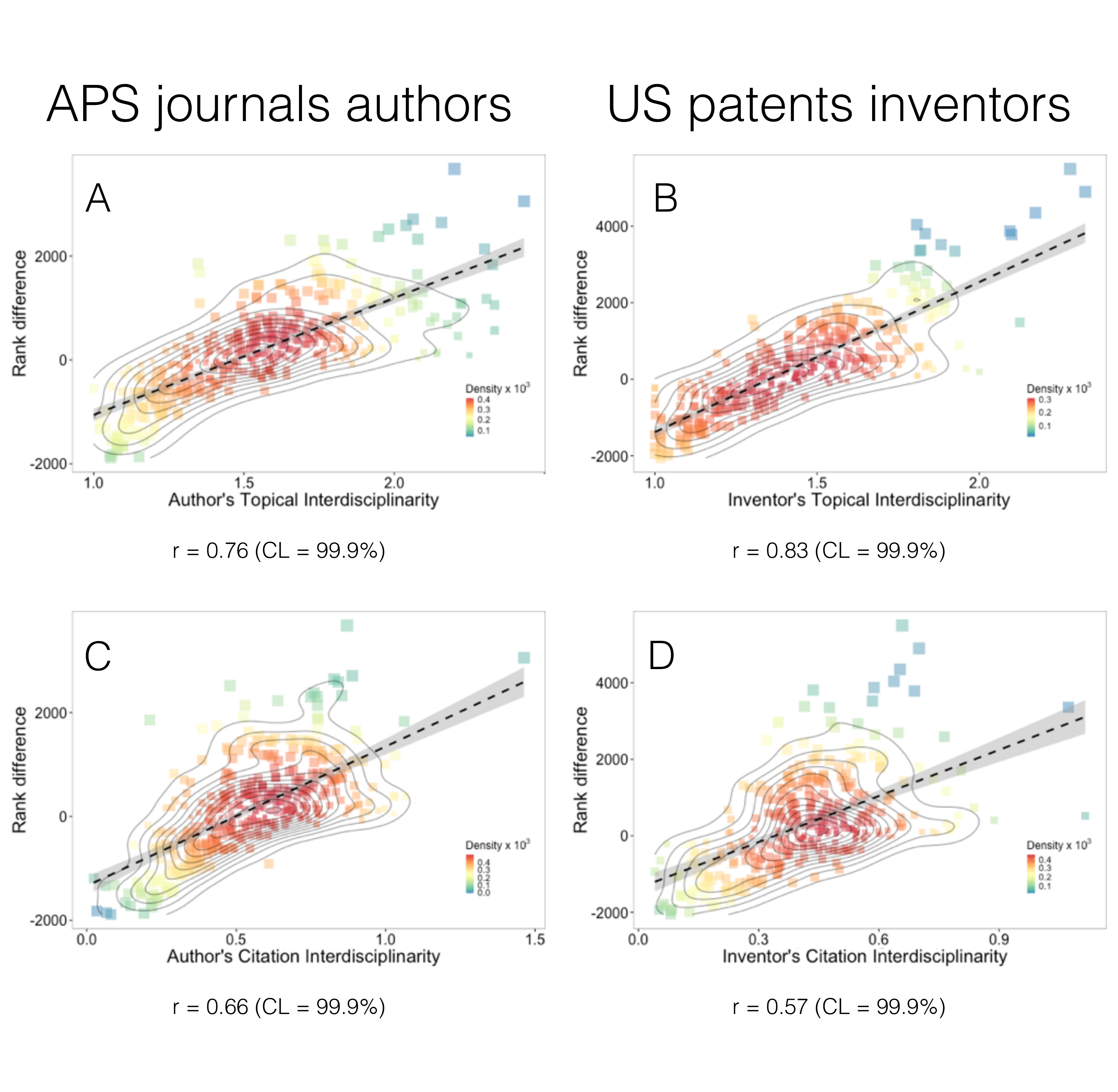}
\caption{\textbf{Correlations.} Heat-maps representing the correlation between the gain in rank that scholars and inventors with $50\pm2$ publications obtain when evaluated using the proposed method -- instead of a method based on a flat representation of the citation network --, and two measures of their interdisciplinarity level. The x-axis represents, in panel (a) and (b), scholars' and inventors' topical interdisciplinarity, defined as the average number of different scientific areas their publications pertain to, and, in panel (c) and (d), their diversity in terms of disciplines of the scholars' and inventors' incoming citations (citation interdisciplinarity). Correlations are calculated using Pearson's $r$ coefficient, and setting the statistical significance at $0.1\%$. Solid lines represent density gradient contours, and dashed lines represent linear regression models estimated via maximum-likelihood.}
\label{control_50papers}
\end{figure}

{\color{black} In Figure~3 of the main text, we show that we find a strong positive correlation between the gain in rank that scholars and inventors obtain when evaluated using the proposed method -- instead of a method based on a flat representation of the citation network --, and their topical interdisciplinarity (panels (a) and (b)). Moreover, we show that the rank gain is also positively correlated with the disciplinary diversity of scholars' and inventors' incoming citations (panels (c) and (d)). To control for the effects of productivity, i.e. the fact that a researcher that has produces more papers has more chances to publishes in more areas or to be cited by papers in many different areas, we perform the same analysis on two subsets of the data by considering in each case only authors with a fixed number of publications. Figure~\ref{control_20papers} shows the result for the subset of authors and inventors with $20\pm2$ publications, and Figure~\ref{control_50papers} for authors and inventors with $50\pm2$ publications. The results are consistent with those obtained using the whole dataset.}

\addcontentsline{toc}{section}{References}
\begin{small}
\bibliographystyle{apsrev4-1} 

\end{small}

\end{document}